%% file: main.tex
\documentclass[sigchi]{acmart}

\usepackage{balance}
\usepackage{graphics} 
\usepackage{booktabs}
\usepackage{xcolor,colortbl}
\usepackage{comment}

\usepackage{tabularx}
\usepackage{tipa}
\usepackage{enumitem}
\usepackage{url}

\usepackage{breakurl}
\usepackage[font=small]{caption}
\usepackage{enumitem}
\setlist{itemsep=0pt}
\usepackage{soul}
\usepackage{xspace}
\usepackage{microtype}
\usepackage{balance}

\usepackage{todonotes}

\title[``There's so much responsibility on users right now:'' Expert Advice for Staying Safer From Hate and Harassment]{``There's so much responsibility on users right now:'' \\Expert Advice for Staying Safer From Hate and Harassment}

\copyrightyear{2023}
\acmYear{2023}
\setcopyright{rightsretained}
\acmConference[CHI '23]{Proceedings of the 2023 CHI Conference on Human Factors in Computing Systems}{April 23--28, 2023}{Hamburg, Germany}
\acmBooktitle{Proceedings of the 2023 CHI Conference on Human Factors in Computing Systems (CHI '23), April 23--28, 2023, Hamburg, Germany}\acmDOI{10.1145/3544548.3581229}
\acmISBN{978-1-4503-9421-5/23/04}

\keywords{Security and privacy, hate, harassment, advice}

\ccsdesc[500]{Security and privacy~Human and societal aspects of security and privacy}

\author{\vspace{0pt}Miranda Wei}
\email{weimf@cs.washington.edu}
\affiliation{%
  \institution{University of Washington, Google}\country{USA}
}

\author{\vspace{0pt}Sunny Consolvo}
\email{sconsolvo@google.com}
\affiliation{%
  \institution{Google}\country{USA}
}

\author{\vspace{0pt}Patrick Gage Kelley}
\email{patrickgage@acm.org}
\affiliation{%
  \institution{Google}\country{USA}
}

\author{\vspace{0pt}Tadayoshi Kohno}
\email{yoshi@cs.washington.edu}
\affiliation{%
  \institution{University of Washington}\country{USA}
}

\author{\vspace{0pt}Franziska Roesner}
\email{franzi@cs.washington.edu}
\affiliation{%
  \institution{University of Washington}\country{USA}
}

\author{\vspace{0pt}Kurt Thomas}
\email{kurtthomas@google.com}
\affiliation{%
  \institution{Google}\country{USA}\vspace{.5cm}
}

\renewcommand{\authors}{Miranda Wei, Sunny Consolvo, Patrick Gage Kelley, Tadayoshi Kohno, Franziska Roesner, Kurt Thomas}

\begin{document}

\input{macros}
\input{00_abstract}
\maketitle

\input{01_intro}
\input{02_related}
\input{03_methods}
\input{04_results_threats}
\input{05_results_advice}
\input{06_results_strategies}

\input{07_discussion}

\input{08_conclusion}
\input{acks}

\bibliographystyle{ACM-Reference-Format}
\balance
\input{main.bbl}

\end{document}

%% file: macros.tex
\renewcommand{\paragraph}[1]{\vspace{5pt}\noindent\textbf{#1.}\xspace}
\newcolumntype{R}{>{\raggedleft\arraybackslash}X}
\newcolumntype{L}{>{\raggedright\arraybackslash}X}
\newcolumntype{P}[1]{>{\centering\arraybackslash}p{#1}}

\newcommand{\eg}{e.g., }
\newcommand{\etc}{etc.}
\newcommand{\ie}{i.e., }
\newcommand{\etal}{et al.\@\xspace}


\newcommand{\numexperts}{24\xspace}
\newcommand{\numadvice}{45\xspace}
\newcommand{\numpostadvice}{6\xspace}

\newcommand{\inlinequote}[2]{``#2''\xspace}
\newcommand{\pquote}[2]{\begin{quote}\textit{``#2''} --\,#1\end{quote}}


\newcommand{\advicetwofa}{\textit{enable any form of 2FA for your most important accounts}\xspace}
\newcommand{\advicepin}{\textit{use a strong PIN or passcode for your devices}\xspace}
\newcommand{\advicestrongpwd}{\textit{use a strong, unique password for all of your accounts}\xspace}
\newcommand{\advicepwdmanager}{\textit{use a password manager}\xspace}
\newcommand{\advicecellpin}{\textit{have a PIN or verbal password associated with your cell provider}\xspace}
\newcommand{\advicesecquestion}{\textit{if a website uses security questions ... use a password-like response}\xspace}
\newcommand{\adviceunlinkableaccts}{\textit{create a pseudonym or use a different email for each of your online accounts}\xspace}

\newcommand{\advicelimit}{\textit{limit sharing of personal information online generally, being conscious of incidental information leaks}\xspace}
\newcommand{\adviceminpii}{\textit{use 3rd-party services to help minimize personal information available online}\xspace}
\newcommand{\adviceperiodicallydel}{\textit{periodically delete old social media posts, messages, and emails}\xspace}
\newcommand{\adviceneversend}{\textit{never send intimate images}\xspace}
\newcommand{\adviceencryptintimate}{\textit{encrypt and/or keep intimate imagery offline}\xspace}
\newcommand{\advicealerts}{\textit{set up alerts to monitor where your name appears in search results (like Google Alerts)}\xspace}
\newcommand{\advicesharephone}{\textit{never share your personal phone number publicly and/or use a second phone number (like Google Voice)}\xspace}
\newcommand{\adviceshareaddress}{\textit{never share your home address publicly}\xspace}
\newcommand{\advicesecondemail}{\textit{use an second email address when signing up for websites or creating new accounts}\xspace}
\newcommand{\advicepublicrecords}{\textit{ensure all public records like registered domains or housing records are tied to a pseudonym}\xspace}
\newcommand{\advicesetprivacy}{\textit{set restrictive privacy settings on social media (like using a Privacy Check-Up tool)}\xspace}
\newcommand{\advicerequestremoval}{\textit{find your personal information or intimate images in search engines or social media sites to remove or request your data be removed}\xspace}
\newcommand{\advicedigitalids}{\textit{don't keep digital copies of your IDs (like a driver's license or passport)}\xspace}

\newcommand{\advicesecuremsg}{\textit{use secure messaging apps for communication}\xspace}
\newcommand{\advicecamera}{\textit{keep your web camera covered when you aren’t using it}\xspace}
\newcommand{\adviceav}{\textit{use antivirus software to detect spyware on your devices}\xspace}
\newcommand{\advicelocationshare}{\textit{never share location information with apps or in posts or photos, including in photo metadata}\xspace}
\newcommand{\advicegps}{\textit{disable GPS when not needed to prevent location tracking}\xspace}
\newcommand{\advicedelayedpost}{\textit{don’t post photos of your activities until after you've left the location}\xspace}
\newcommand{\adviceuntag}{\textit{ask friends and family not to share posts or photos about you, and untag yourself in any posts or photos that are shared}\xspace}
\newcommand{\adviceappminimize}{\textit{avoid downloading apps you do not need and remove already downloaded apps that you no longer need}\xspace}
\newcommand{\advicepobox}{\textit{use a virtual or PO mail box rather than sharing your home address}\xspace}
\newcommand{\adviceairtags}{\textit{do a physical search or digital scan for tracking devices like Airtags or Tiles}\xspace}
\newcommand{\advicesecondsim}{\textit{use a second, separate SIM card to prevent tracking of your location or phone calls}\xspace}

\newcommand{\advicemoderate}{\textit{use platform-provided tools to automatically filter or moderate abusive messages}\xspace}
\newcommand{\advicemute}{\textit{mute people who post abusive messages}\xspace}
\newcommand{\adviceblock}{\textit{block people who post abusive messages}\xspace}
\newcommand{\adviceselectivecommunity}{\textit{be selective about which online communities you participate in}\xspace}
\newcommand{\adviceselectiveaspect}{\textit{be selective about when and to whom you reveal marginalized aspects of your identity}\xspace}
\newcommand{\adviceleave}{\textit{leave a platform entirely}\xspace}

\newcommand{\advicefamilyimpersonation}{\textit{ask friends, family, and colleagues to help keep an eye out for impersonation}\xspace}
\newcommand{\adviceverifyaccount}{\textit{request for your account to be verified}\xspace}
\newcommand{\adviceaccountsquatting}{\textit{create accounts with your name on all major platforms}\xspace}

\newcommand{\advicevpn}{\textit{use a VPN while online to hide your IP address}\xspace}
\newcommand{\adviceddos}{\textit{get DDoS protection for personal websites}\xspace}

\newcommand{\adviceswat}{\textit{reach out to law enforcement in advance to warn about you being a potential target of swatting}\xspace}

%% file: 00_abstract.tex
\begin{abstract}
Online hate and harassment poses a threat to the digital safety of people globally. In light of this risk, there is a need to equip as many people as possible with advice to stay safer online.
We interviewed 24 experts to understand what threats and advice internet users should prioritize to prevent or mitigate harm. As part of this, we asked experts to evaluate 45 pieces of existing hate-and-harassment-specific digital-safety advice to understand why they felt advice was viable or not. We find that experts frequently had competing perspectives for which threats and advice they would prioritize. We synthesize sources of disagreement, while also highlighting the primary threats and advice where experts concurred. Our results inform immediate efforts to protect users from online hate and harassment, as well as more expansive socio-technical efforts to establish enduring safety.
\end{abstract}

%% file: 01_intro.tex
\section{Introduction}\label{sec:intro}
Online hate and harassment is a threat with pernicious reach, negatively impacting the safety---\eg emotional, sexual, or physical safety---of over 48\% of internet users around the world~\cite{thomas2021:sok}. While certain populations are at higher risk of experiencing targeted attacks---such as creators~\cite{thomas2022creators}, journalists~\cite{chen2020you}, gamers~\cite{kowert2020dark}, survivors of intimate partner abuse~\cite{matthews2017stories, freed2017digital}, and people with marginalized identities~\cite{scheuerman2018cscw, amnesty2018troll, citron2016hate, Aghazadeh2018, duggan2017black}---\textit{anyone} can become a target of online hate and harassment. 
Going online today necessitates that internet users navigate a complex array of technology-mediated hate and harassment, such as toxic content, brigading (coordinated abusive behavior online), non-consensual sharing of intimate imagery, or device-enabled location surveillance~\cite{thomas2021:sok}. 
As such, there is a need to prepare as many people as possible with appropriate knowledge and best practices for staying safer.

Advocates have published a wealth of resources to educate potential targets about protections for online hate and harassment. Non-governmental organizations (NGOs) like PEN America's \textit{Online Harassment Field Manual} helps journalists and others in ``navigating online abuse and tightening digital safety''~\cite{penfieldmanual}. Feminist Frequency's \textit{Speak Up \& Stay Safe(r): A Guide to Protecting Yourself From Online Harassment} is ``designed for women, people of color, trans and genderqueer people, and everyone else whose existing oppressions are made worse by digital violence'' ~\cite{speakup}.
Platforms also publish resources, such as YouTube's \textit{Creator Safety Center}, which helps creators ``make a plan to stay safe online''~\cite{youtubecreatorsafety}.

Advice and its framing ranges from general (\eg broadly applicable) to tailored (\eg highly specialized). 
Existing online advice for staying safer from hate and harassment tends to be tailored, such as for marginalized populations that are commonly targeted, or for common potential threats.
Tailored advice is invaluable for populations that experience disproportionate risks, yet there is also an immense challenge to create and maintain unique advice for numerous disparate groups.
There is comparatively little general advice for staying safer from hate and harassment, though general advice will be increasingly beneficial as more people experience hate and harassment. 
Such advice is a valuable addition to---not a replacement for---tailored advice.
Particularly because many targets of hate and harassment may not predict being targeted or seek out advice, general advice establishes a consistent message for advice-givers to repeat at scale, hopefully reaching people before they experience attacks.

In this work, we explore developing general advice to stay safer from online hate and harassment, that is, advice that is broadly applicable and can be given without additional context about the user. 
We engage leading scholars and advocates to synthesize and evaluate the existing landscape of advice, including to identify frequently repeated advice that is not achievable, and to understand what experts believe would make such advice easier to adopt.
We first gathered 219 disparate pieces of advice from existing guides, deduplicated them into 45 protective practices, and further categorized each by the threat it is intended to address.
We focus on safety advice that can be implemented before hate and harassment occurs---\ie prevention or mitigation---and scope advice narrowly to \textit{proactive practices}. We then conducted interviews with 24 subject matter experts (based primarily in Western countries) who work with people experiencing online hate and harassment to assess three research questions:
\vspace{5pt}
    \paragraph{RQ1: Informing user threat models} 
    Which online hate and harassment threats do experts believe most internet users should prioritize taking action to prevent or mitigate, and why?
    
    \paragraph{RQ2: Prioritizing existing advice} For specific hate and harassment threats, how do experts prioritize existing advice for internet users who might experience them, and why?

    \paragraph{RQ3: Recommending overall safety strategies}
    Assuming they do not have details about users' unique situations and there is no known ongoing attack, what are experts' top recommendations for internet users to stay safer from online hate and harassment?  
\vspace{10pt}

Overall, experts felt that most internet users should focus their safety efforts on three of the seven categories of threats~\cite{thomas2021:sok} we asked about: toxic content, content leakage, and surveillance. For some threats---such as account lockout and control, which didn't make the top three---there was a clear prioritization of advice: use two-factor authentication (2FA), use strong passwords, and to a lesser extent, use a password manager. Conversely, expert perspectives on how to mitigate content leakage or surveillance were far more discordant. Advice such as keep your camera covered, use anti-virus to detect spyware, or never share your location information with apps drew a range of perspectives.
Towards overall safety strategies for minimizing harm, we find that experts recommended a mindset of data minimization, staying abreast of classic security advice, being self-aware and self-determined online, as well as participating in and fostering healthier online communities.

Our findings underscore a reality echoed by nearly every expert we spoke with: safety from online hate and harassment currently falls predominantly on users to enact. 
Experts judged that alleviating this burden would require pro-social, community-building approaches to increase safety for all. 
For advocates designing education materials, our work exposes the current state of generally applicable advice as well as multiple competing priorities that need to be considered when creating and delivering advice. For platform developers, our analysis surfaces gaps in protections and limitations of existing safety tools that lead to experts not recommending their use. And finally for users, our research provides a ranking of the most impactful existing advice that can be enacted today.

%% file: 02_related.tex
\section{Related Work}\label{sec:related}

\paragraph{Experiences of Hate and Harassment}
Hundreds of millions of people globally experience online hate and harassment~\cite{thomas2021:sok, microsoft2019dci, pew2017harassment}, enduring serious physical, emotional, professional, relational, and financial harms~\cite{citron2016hate, scheuerman2018cscw}. Prior research into online hate and harassment and protective practices is expansive. We rely on a taxonomy of experiences from Thomas \etal~\cite{thomas2021:sok}, which synthesizes the literature into seven categories of threats: toxic content (\eg bullying, hatespeech, trolling),
content leakage (\eg doxxing, non-consensual intimate images),
overloading (\eg brigading, dogpiling, denial of service), 
surveillence (\eg stalking), false reporting, impersonation, and lockout and control (\eg account takeover).

Online hate and harassment often builds on other axes of oppression. Harm tends to be disproportionately experienced by marginalized people, \eg transgender people~\cite{scheuerman2018cscw}, women~\cite{citron2016hate, jeong2018garbage, sambasivan2019they, vitak2017identifying, amnesty2018troll, im2022perspectives}, and Black and other marginalized racial or ethnic groups~\cite{jeong2018garbage, duggan2017black, amnesty2018troll}. Attacks are more likely to be perpetuated by privileged groups such as men with a greater social dominance orientation~\cite{tang2016mens}. Attacks may also narrowly target at-risk users in an attempt to silence voices---such as journalists, gamers, and creators~\cite{kowert2020dark, chen2020you, thomas2022creators, snyder2017fifteen}---or coerce and control individuals as in intimate partner abuse~\cite{freed2017digital, freed2018stalker, matthews2017stories}.
The broad reach of online hate and harassment, and the reality that many individuals are unaware of the risks until they experience an attack, underscores the need to provide generally applicable advice for staying safe as a precursor to tailored advice.

\paragraph{Providing General Security Advice}
Security advice should be effective, actionable, and understandable~\cite{redmiles2020comprehensive}, as well as consistent and concise~\cite{berdan2021journalists}.
Unfortunately, the collective state of security advice (not just for online hate and harassment) is far from concise, with experts offering hundreds of pieces of advice~\cite{reeder2017simple, redmiles2020comprehensive}. 
Fragmentation means that users learn advice from different sources~\cite{redmiles2017divide}---including stories~\cite{rader2012stories, rader2022stories} or social ``triggers''~\cite{das2019triggers}---depending on skill levels and socioeconomic status~\cite{redmiles2016learned}, age~\cite{nicholson2019infoseeking}, or other factors. Claims that advice is helpful are easy to make, but empirically impossible to refute~\cite{herley2016unfalsifiability}, leading many researchers to call for prioritization~\cite{reeder2017simple, berdan2021journalists, herley2014more, ion2015noone}. Security advice is often perceived to offer a poor cost-benefit tradeoff---high cost, low benefit---so motivation to follow advice is weak~\cite{herley2009externalities, fagan2016motivates}.
To aid adoption, the delivery of advice should help people understand why the advice would benefit them~\cite{berdan2021journalists, herley2014more, herley2016unfalsifiability}. We explore themes related to prioritization, cost tradeoffs, and delivery as part of our analysis of advice for staying safer from online hate and harassment.

\paragraph{Tailoring Security Advice} Significant research has also explored how to tailor support and security advice to at-risk groups, such as civil rights protestors~\cite{boyd2021blm, wade2021recommendations}, employees~\cite{dangpham2016climate, dangpham2017share}, human trafficking survivors~\cite{chen2019trafficking}, journalists~\cite{berdan2021journalists, mcgregor2015journalists, mcgregor2016journalists, mcgregor2016obscurity}, older adults~\cite{nicholson2019infoseeking}, politicians~\cite{consolvo2021campaigns}, queer individuals~\cite{geeng2022lgbtq}, refugees~\cite{simko2018computer}, and sex workers~\cite{mcdonald2021sexworkers}.
In tailoring advice for specific populations, these studies lie on the opposite end of a spectrum from the studies of general security advice described earlier.
Advice could also be tailored by specific hate and harassment threats, but little academic research seems to have used that lens.

Though specialization enables more targeted support to groups that have been historically overlooked, it also enshrines criteria for additional support, \ie group membership.
For some groups, membership is evident or persistent (\eg by identity, career), but may not be for the ever-increasing set of people who experience online hate and harassment. 
Some potential targets may not seek out tailored advice or even realize that they are at risk until after an attack is underway.
Further, as the number of groups increases, creating and maintaining unique tailored advice becomes progressively difficult.
To grapple with such difficulties, this work explores developing general advice, absent specific user information.

\paragraph{Platform Safety Affordances}
Almost all major platforms that allow user-generated content now explicitly prohibit hate and harassment~\cite{pater2016characterizations}, and they are continually building features to combat online hate and harassment. Automated features to reduce online hate and harassment include automated moderation of content~\cite{chandrasekharan2017bag, singhal2022sok, jennifer2022feels, kim2021human, razi2021sexual, bursztein2019rethinking} or accounts~\cite{im2020still, saeed2022trollmagnifier, ribeiro2018characterizing}.
In terms of manual efforts, platforms allow individuals~\cite{crawford2016flag} or authorized reporters~\cite{matias2015reporting} to report offending content (although the subsequent decisions can be seen as unfair or opaque~\cite{pan2022legitimacy}) or to implement crowdsourced blocklists~\cite{jhaver2018online, geiger2016blockbots}.
In particular situations, users or communities that are determined to be harmful have been  deplatformed entirely~\cite{chandrasekharan2017you, horta2021migrations, ali2021understanding}.
Other efforts aim to provide peer support for users experiencing hate and harassment (\eg Squadbox for email~\cite{mahar2018squadbox} and the Heartmob support community~\cite{heartmob, blackwell2017classification}).
In our work, we investigate experts' opinions of the current state of safety online, noting when they support advice recommending certain affordances, or when none exist to protect against certain attacks.

%% file: 03_methods.tex
\section{Methods}\label{sec:methodology}
We interviewed \numexperts hate and harassment subject matter experts in July and August 2022 to discuss what advice might be generally applicable, that is, they would give to ``general internet users'' to stay safer from online hate and harassment. We use to this term throughout the remainder of this paper to capture most internet users, irrespective of their risk level, as anyone can be targeted by online hate and harassment. 
As part of this, we also explored the complexities of providing safety advice in a general manner (\ie not targeted to particular groups) and how to prioritize a large body of safety advice for users with limited time and resources. 

\subsection{Recruiting \& Participants}
We recruited subject matter experts---hereafter referred to as \textit{experts}---who had a background in providing support to people experiencing
online hate and harassment.
Towards developing advice that would be general and widely applicable, we aimed to recruit participants who represented a diverse set of roles, populations assisted, and geographies.
We made sure to recruit experts who had experience supporting marginalized populations.
We identified 55 experts and organizations involved in the development of the advice guides we gathered (see Section \ref{ss:gathering-advice}), had publications related to hate and harassment safety practices, or were professional contacts. We directly solicited their participation via email; 24 participated in our study.
Our 24 participants were academics\footnote{Participants' academic departments included Computer Science, Journalism, Information Sciences, Public Policy, Criminology, and Human-Computer Interaction.}  (n=12), NGO employees (7), and industry professionals (6).\footnote{Totals do not add up to \numexperts due to multiple roles.} Their specializations included social media (7), gaming (6), journalism (4), intimate partner abuse (3), online content creators (2), youth (1), activists (1), and attacker coordination (1). Participants' had two to 40 years of experience (average: 10 years, total: 237 years) in roles related to hate and harassment. Participants primarily operated in the U.S.\ (20), but also the U.K.\ (2), Australia (1), and Turkey (1), additionally speaking about France (1) and the Caribbean (1). We caution that no set of experts can comprehensively cover all people who experience online hate and harassment (\eg all demographics, all occupations). We discuss this limitation further in Section \ref{ss:limitations}.

\subsection{Gathering Advice}\label{ss:gathering-advice}
Prior to conducting the interviews, we aggregated existing digital-safety advice related to hate and harassment. We gathered the advice from online searches, preliminary discussions with experts, and the domain knowledge of the authors of this work. 
We collected 49 online support resources, then filtered out those that did not address proactive practices (27), did not provide actionable advice (8), or only incidentally addressed hate and harassment (6).\footnote{The complete list of advice guides that informed our work is included in the supplementary material.}
Resources targeted audiences such as general internet users (\eg OnlineSOS, Consumer Reports), social media users (\eg Heartmob), journalists (\eg PEN America), youth (\eg Planned Parenthood), and more. 
Of the final set of 15 resources, five were tailored to specific at-risk populations, five to specific threats, and three to specific at-risk populations facing specific threats. Only two were not tailored (\ie for anyone who might face hate and harassment online).

Across the support resources were 219 pieces of non-unique advice. Two researchers engaged in affinity diagramming to deduplicate advice and identify which of the seven categories of hate and harassment the advice best helped prevent or mitigate~\cite{thomas2021:sok}. This effort resulted in 45 unique pieces of advice. As part of this process, we omitted advice about ongoing attacks (\eg ``deactivate accounts if you are being doxxed'') or recovery, as our focus was on proactive practices.

As part of our interview protocol, we asked participants whether there was any additional advice they felt was missing. 
After applying the same scoping criteria as before and deduplicating advice, participants identified six ``new'' pieces of advice in total, demonstrating our approach achieved sufficient coverage of most advice. Of those six pieces, only one was mentioned by more than two experts. We discuss new advice in Section~\ref{ss:advice-ranking}.

\subsection{Study Procedures \& Data Collected} \label{ss:procedures}
Our semi-structured interview protocol consisted of four phases that were completed in a single, remote session with each participant.\footnote{Our interview script is included in the supplementary material.}
First, we asked participants about their background in helping to protect people from online hate and harassment, as well as any specific populations they assisted. 

Second, we asked participants to rank which of the seven categories of hate and harassment threats general internet users should prioritize preventing or mitigating~\cite{thomas2021:sok}.
Given prior work emphasizing the need for minimalism and prioritization~\cite{redmiles2020comprehensive, consolvo2021campaigns}, we developed this activity to require a discrete ordering. We asked experts to ``think aloud''~\cite{charters2023thinkaloud} while ranking to capture their underlying thought processes and opinions on each threat category.

Third, participants engaged in a card sorting activity, continuing to ``think aloud,'' where they categorized the \numadvice pieces of advice into ``High,'' ``Medium,'' or ``Low'' priority, or advice they ``Don't recommend.''
Rather than sorting all \numadvice pieces at once, this phase was broken into five parts, based on the seven categories of threats that each piece of advice was best positioned to prevent or mitigate.\footnote{In the event an expert felt a piece of advice spanned multiple threats, we discussed with experts what implications that had for the advice and its priority to capture any missed nuance.} 
The 5 parts were: 

\begin{enumerate}
    \item Lockout \& Control -- \textit{9 pieces of advice to sort}, 
    \item Content Leakage -- \textit{13 pieces}, 
    \item Surveillance -- \textit{11 pieces}, 
    \item Toxic Content -- \textit{6 pieces}, and
    \item Impersonation, Overloading, \& False Reporting -- \textit{6 pieces}.
\end{enumerate}
We decided on this approach during pilot testing. We found that it helped participants avoid over-indexing on the threat (which we captured in the second phase), and instead focus on the task of ranking individual pieces of advice. This partitioning also reduced the cognitive load of comparing \numadvice pieces of advice at once. 
After participants had sorted all advice in one threat category and if it had not yet been mentioned, we asked participants what, if any, advice was missing for that threat.

Lastly, we asked participants to enumerate the top three overall recommendations they would give to a general internet user to stay safer from online hate and harassment (which could be independent of the advice they ranked).
We then engaged in an open discussion about the challenges of delivering advice; what, if any, existing advice guides they thought were effective; and ecosystem changes that might help shift the burden of staying safer from online hate and harassment away from users.

All interviews were led by the same researcher. They lasted from 63 to 97 minutes (average: 88 minutes). Each participant received a \$100 USD gift card (or equivalent local currency) as a thank you. The amount was set by our institution for studies involving experts.

\subsection{Analysis Approach}
We used a mixed quantitative and qualitative approach to analyze our data, informed by how our knowledge and expertise is situated~\cite{haraway1988situated}.
Our team's primary lens is security and privacy, with additional expertise in social media, online safety, and human-computer interaction.
Our research and analysis focused on technical advice that users could follow to stay safer from online hate and harassment, which is only one of many approaches to digital safety.

From our semi-structured interviews, we gathered ordinal and count data about how experts ranked threats and pieces of advice.
We quantitatively analyzed this data to produce an average ranking of the threats (RQ1) and proportions of how experts prioritized the advice (RQ2), as well as to inform the order of results subsections. 
To add qualitative depth, we applied thematic analysis to experts' open-ended responses to understand the factors that informed their threat prioritization (RQ1) and advice evaluation (RQ2), as well as generate themes from experts' top safety strategies (RQ3).
We use thematic analysis~\cite{braun2022conceptual}, both inductively and deductively, because of its flexibility with respect to theory or goal, and its emphasis of researcher subjectivity as ``analytic resource'' for interpretation~\cite{braun2021onesize}.
With a deductive approach, we referred to our own domain knowledge, as well as prior work, to direct our analysis of which factors informing threat prioritization and advice evaluation we thought might be relevant (\eg severity and agency~\cite{scheuerman2021framework}, effectiveness and actionability~\cite{redmiles2020comprehensive}).
To analyze factors that experts talked about as important, we used an inductive approach~\cite{thomas2006:inductive}, and focused on the semantic (\ie reflecting what experts explicitly said) as opposed to latent (\ie experts' underlying assumptions)~\cite{braun2022conceptual}.

During interviews, a researcher who was not leading the interview took notes, focusing on capturing content.
For analysis, notes were reformatted from per-interview to per-research question, \ie threat ranking, advice prioritization, and overall top advice.
One researcher read and re-read all responses, and developed a list of rationales (\ie themes) that participants used to prioritize threats and evaluate advice, as well as categories of participants' top advice. 
We reviewed our ideas by revisiting the data, writing reflective memos, regularly meeting with members of the team, and iteratively updating the themes until we felt we had reached meaning sufficiency~\cite{braun2022conceptual}.
In the results, we report quotes (transcribed from interview recordings) to illuminate (a) instances where experts largely agreed, and/or (b) nuances on which experts disagreed, but were novel and insightful.

\subsection{Ethics}
Our study plan was reviewed by experts at our institution\footnote{This study was conducted at Google.} in domains including ethics, human subjects research, policy, legal, security, privacy, and anti-abuse. We note that our institution does not require IRB approval, though we adhere to similarly strict standards.
Prior to any data collection, all participants signed a consent form, which included agreement to record their session. At the start of each session, we re-confirmed consent (two participants requested that their sessions not be recorded, so they turned off their cameras and we only recorded audio and screens for the card sorting with their permission). We also reminded participants that their engagement was entirely voluntary; they could pause, skip activities, or stop the session at any time and still receive the full thank you gift. 

We protected our study data---including videos, audio, notes, and transcripts---by encrypting all records at-rest, restricting access to only the core research team (and institutional administrators), and requiring two-factor authentication with a physical security key to access the information. Video recordings, audio recordings, and transcripts were set to auto-delete after 6 months, though we kept some anonymized notes to be used in the publication process. Finally, we asked each participant whether they would like to be recognized in any acknowledgements or materials produced as part of the research. As a best practice, we attribute quotes only to a participant ID; we specifically omit unique details, phrases, or words from quotes to mitigate identification of participants.

\subsection{Limitations} \label{ss:limitations}
Given the breadth of digital-safety experiences, our evaluation of advice is non-exhaustive and limited to the \numadvice pieces of advice we identified prior to our study, and the \numpostadvice additional pieces of advice mentioned by participants. Our de-duplication of hundreds of pieces of similar advice may have resulted in omitting nuanced language that some experts viewed as important to the delivery. Many participants viewed advice through the lens of the populations they help protect (\eg gamers, journalists, \etc), as well as through their geographic biases, highlighting the challenges of generalized safety advice in the absence of additional information about the person seeking help.
Nevertheless, we reached meaning sufficiency~\cite{braun2022conceptual} on the themes for how experts prioritized threats and evaluated advice before concluding our final interview. 

General advice, compared to tailored advice, is unavoidably less accurate and thus might consider the wrong threats for some individuals. General advice might have limited benefits for those experiencing extreme instances of hate and harassment and unnecessary costs for those who do not experience any. We were interested in exploring this limitation of general advice, so we asked experts how they would rank potential threats for a general audience. We report their rankings and thought processes in our results.

Relatedly, our use of the term ``general internet user'' in interviews may have introduced biases; most of our experts were in the U.S. where white men are assumed to be the default persona~\cite{morris2016standard}. 
To combat these biases, we recruited experts with a range of perspectives and backgrounds, and also asked experts to explain who they imagined advice would or would not serve.

%% file: 04_results_threats.tex
\section{Results}\label{sec:results}

Most experts agreed on three categories of hate and harassment threats that general users should prioritize taking action to prevent or mitigate: toxic content, content leakage, and surveillance. 
Experts commonly used three dimensions---severity, prevalence, and agency---as ranking criteria for evaluating the seven categories of threats (Section~\ref{ss:threat-ranking}).
Of the 45 pieces of advice experts were asked to rank, they 
most highly prioritized enabling two-factor authentication (Section~\ref{ss:advice-ranking}). When ranking individual pieces of advice, experts weighed factors such as efficacy, ease of implementation, and effect on online participation. 
Experts' top overall advice recommended minimizing personal data online and developing an awareness of the unique threats that one might be targeted by, as well as taking pro-social actions to build safer online communities (Section~\ref{ss:advice-top}). In this section, we discuss each of these findings in further detail. 

\begin{table}[t]
    \centering
    \input{tables/top_threats}
    \caption{Ranking of hate and harassment threats. This includes overall average ranking (highest = 1, lowest = 7), the number of experts who ranked a threat as a top priority (maximum of 24), and the number of times experts ranked a threat as one of their top three priorities (maximum of 24).} 
    \Description{Table containing rankings from experts on the top hate and harassment threats.}
    \label{table:threat_ranking}
\end{table}

\subsection{Ranking Potential Threats}\label{ss:threat-ranking}
As part of the study, experts ranked which, if any, of seven categories of hate \& harassment-related threats internet users should prioritize protecting themselves from, and why. In this section, we describe the criteria experts used to rank the categories, then review results for each category.

\paragraph{Ranking criteria}
As shown in Table~\ref{table:threat_ranking}, experts were split on the foremost category of threat they thought internet users should prioritize. 
This was, in part, due to differences in the criteria 22 of our 24 experts used while ranking (two did not mention any criteria). Their ranking criteria included the \textit{severity} of (potential) harms that might result from a threat, the \textit{prevalence} of the threat (\ie the likelihood of an attack occurring), and the \textit{agency} of users to mitigate the threat. 

For 10 experts, \textit{severity} of (potential) harms was their primary criterion when ranking threats, and particularly threats to ``physical safety, their bodily integrity, [as well as] to their mental health'' (P22), echoing Scheuerman \etal's \textit{Framework of Severity}~\cite{scheuerman2018cscw}. 
One expert favored this strategy because it allocated attention to those most in need of help:
\pquote{P21}{People who are targeted by the most severe forms of online hate and harassment are in marginalized communities and they need additional protections.}

Nine experts relied on \textit{prevalence} as their primary criterion for ranking threats. Experts expressed that this meant any guidance would better resonate with internet users, as it reflected attacks they were more likely to encounter.  As P18 explained: ``What is the most prevalent problem right now... that people need to be aware of?'' For other participants, prevalence reflected a disciplinary norm that stemmed from limited time and resources:
\pquote{P1}{In computer security, you want to educate people about attacks or threats they are likely to encounter. There are some attacks that are only relevant to government agencies, or high-profile organizations and so on.}

Three experts used \textit{agency} as their primary criterion for ranking. These experts remarked on the importance of building on user self-efficacy: ``What is the lightest lift for a user?'' (P23). These experts focused on which threats had the most meaningful existing protections, or where ``a well-timed warning or educational intervention'' (P20) might be effective.

The differences across our experts in the primary criterion---and even secondary and tertiary criteria---they used for ranking emphasize a challenge for protecting internet users from hate and harassment: there is no consensus yet for which problems to prioritize, or even \textit{how} to prioritize them. While rankings may meaningfully differ for at-risk groups, many members of those groups may be unaware they are at-risk, or an event may suddenly put them at-risk~\cite{warford2021sok}. General awareness of certain hate and harassment threats can thus provide critical, early protection before they are targeted.
In this light, we explore which threats stood out more than others for experts, and where opinions diverged.

\paragraph{Toxic content} On average, toxic content---which includes bullying, hate speech, and sexual harassment---ranked as the highest priority threat across experts, often because of its prevalence.
P15 noted that it was ``the number one type of harassment that I see.'' 
Others added that toxic content could incur emotional harm and have ``significant long-term repercussions'' (P16), and that some users ``might not even know that they are [experiencing it]'' (P6), contributing to a greater need for users to prioritize learning what constitutes toxic content and taking proactive measures to prevent it.

Some experts ranked toxic content with lower priority, as---though it can cause harm---it ``usually doesn't get to physical, severe harm'' (P13) and because prevention is better handled at the community-level: ``toxic content normalizes certain types of behavior, so it's a greater danger as a community norm than towards an individual'' (P19). Others ranked it lower priority, saying that users had more agency: 

\pquote{P3}{You can remove yourself from those situations either by logging out or by initiating or installing all of the protection features that a lot of online platforms have. It really sucks... [sending toxic content] is not okay---no one should do that---but you can remove yourself from those situations.}

\paragraph{Content leakage}
Content leakage---which includes doxxing and non-consensual sharing of intimate images---was ranked the second highest threat on average. Experts pointed to how common this threat is---``people send sexts all the time'' (P10)---though often underestimated the risks, because people ``really cannot imagine what it's like to be doxxed'' (P21). The severity of content leakage, experts judged, arose because leakage is irreversible and attacks could easily spill over into users' ``real lives, their experience of life outside'' (P3) such as by facilitating stalking. 
Conversely, other experts rated content leakage a lower priority because it is less prevalent---``requires more work from the trolls'' (P4)---or because users have less agency to prevent it:

\pquote{P24}{I can't think of any particular platform that really does an effective job of full control of [content leakage]... A lot of people have to escalate. So it's not just primarily relying on tools in the online space, but looking at resources that could help them seek justice offline.} 

\paragraph{Surveillance}
Just five experts ranked surveillance---which includes stalking and monitoring accounts or devices---as the foremost threat in the context of hate and harassment, though it featured in 12 experts' top three. In general, experts felt surveillance was unlikely to be prevalent and was ``more context dependent'' (P19). Though experts noted that it had the potential to cause severe harm (\eg it can be a ``high risk to physical safety''), P22 thought that people had more agency to prevent it (\ie people ``generally have more control and can find technical solutions''). 

Experts emphasized three contexts where this prioritization changed. The first was individuals experiencing intimate partner abuse, as surveillance ``often begins before people realize they're in an abusive relationship'' (P12), preceding the phases of abuse as identified in Matthews \etal~\cite{matthews2017stories}. The second was for people in civil society targeted by government-backed harassment and trolls: ``one of the biggest digital issues [for journalists], [it] leads to physical threats and imprisonment, or assassination'' (P4), and third, for prominent individuals~\cite{warford2021sok} as attacks were ``more relevant for popular accounts for people of a certain reputation'' (P1).
Experts broadly commented that incidents with surveillance could be exceptionally severe for targets:

\pquote{P24}{It's one of those thing where if it happens to you, it's going to have a significant impact emotionally and for your physical safety. In terms of long term consequences, it impacts how you interact in online spaces.}

\paragraph{Lockout and control}
Experts disagreed on how prevalent lockout and control---manipulating devices, being maliciously locked out of one's account---would be for an internet user specifically in the context of online hate and harassment.

However, many felt this was a more general security threat due to the prevalence of phishing and data breaches.
For example, P8 noted that the ``prevalence is high if you're vulnerable to a credential stuffing attack'' while P17 ranked this threat the lowest because it is ``not a primary way perpetrators attack people in the context of hate and harassment.''

Regardless of the prevalence of this threat, experts remarked that being locked out of accounts and devices could facilitate other threats. Experts emphasized that targets  ``have to lock down [their] accounts and personal information first'' (P14) in order to prevent down-stream harms, such as content leakage or surveillance. In this way, experts prioritized account security as a locus of agency: 

\pquote{P3}{[Lockout and control] strikes me as the most invasive. So anything where somebody feels like they don't have control over their own content to me, is the number one [priority].}

\paragraph{Impersonation}
Only one expert ranked impersonation---fake profiles or communication posing as the target---as their foremost threat, commenting that it poses a ``very immediate threat to personal information, devices, and can have a very large effect on someone's life'' (P14). In terms of severity, experts agreed about the potential for impersonation to affect an individual's emotional well-being and reputation, as well as ``collective harm on people in your network'' (P24). Similar to surveillance, experts noted the low prevalence for most internet users, though it could be higher priority for prominent figures.

Impersonation was seen as harder to prepare for, or even not preventable at all.
One expert pointed out the precarity of people who have begun to gain public followings, but may not have all the resources of more prominent public figures:
\pquote{P21}{The place I see impersonation happen a lot is with low-level influencers... they're less likely to know it; they won't have a [support] team.}

Some experts spoke to the challenges of recovering from impersonation: that marginalized people are harmed the most because there are ``not a lot of tools or legal protections'' (P19) for them, and that it was a ``pain in the butt to get platforms to respond to impersonation reports and get them taken down'' (P23).
One expert with personal experience assisting targets of harassment seemed more optimistic about recovery, saying that in their experience, it ``usually turns out more alright than other situations'' (P10).

\paragraph{False reporting}
No expert in our study ranked false reporting---such as swatting or false abusive account reporting---as the top threat for internet users, though seven put it in their top three. Experts viewed false reporting as a very rare occurrence, though they noted that it was more common on gaming platforms and among ``big armies of trolls'' used by ``authoritarian regimes'' (P4).

Experts noted the severity of harms stemming from false reporting could be extremely divergent or unpredictable. P6 shared that false reporting was a ``standard bullying tactic'' employed by kids---one that might not lead to consequences for those employing it or to those targeted by it (though it would slow triaging legitimate complaints). 
On the other hand, P20 spoke about how swatting could cause extremely severe harm, including being fatal. 
The viability of false reporting as a tactic, and thus agency of users to act, largely fell to the review process of the emergency service or platform contacted, which could be complicated by limited resources:

\pquote{P15}{The claim is usually that the content they have, the video they've shared, or the post is of a `sexual nature.' And it doesn't contain any of it. But because it's in a foreign language that isn't supported by the platform, it's taken down immediately.}

\paragraph{Overloading}
Just three experts ranked overloading---including brigading, notification bombing, or denial of service attacks---in their top three threats; similar to false reporting, none ranked it as the top threat. Most experts commented that while overloading could be frustrating, it has a low prevalence of occurring for most internet users (notable exceptions are those with high profile accounts or websites). For notification-based or network-based attacks, experts felt such attacks were low severity: ``it's not necessarily going to affect your psyche or your personal well-being'' (P4) and ``annoying but not as important'' (P5). Experts expressed that overwhelming volumes of potentially toxic comments could be far more severe:

\pquote{P19}{For an individual to get piled on... that was one of the primary tools that Gamergate used to harm their targets. It was very harmful, the scale of the harm, in addition to the toxicity.}

%% file: tables/top_threats.tex
\begin{tabularx}{\columnwidth}{Xrrr}
\toprule
\bf Threat & \bf Average & \bf  Top & \bf Top 3\\
\bf  category & \bf ranking & \bf threat & \bf threats\\
\midrule
     Toxic Content &  2.88 &         8 &              16 \\
   Content Leakage &  2.92 &         7 &              14 \\
      Surveillance &  3.33 &         5 &              12 \\
 Lockout \& Control &  3.96 &         3 &              12 \\
     Impersonation &  4.25 &         1 &               8 \\
   False Reporting &  4.96 &         0 &               7 \\
       Overloading &  5.71 &         0 &               3 \\
\bottomrule
\end{tabularx}

%% file: 05_results_advice.tex
\subsection{Prioritizing Current Advice}\label{ss:advice-ranking}
\begin{figure*}[t]
\centering
\includegraphics[width=\textwidth]{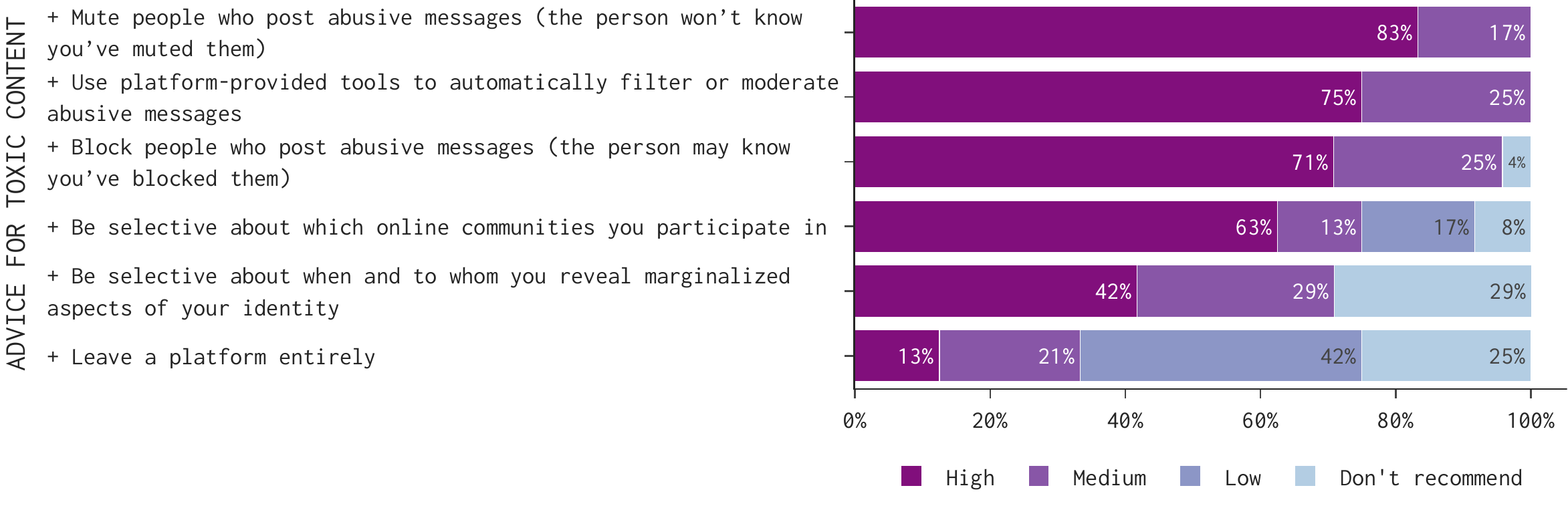}
    \caption{Ranking of advice that users could employ to help prevent toxic content. Experts favored all forms of platform-provided moderation tools over advice that curtailed online participation.}
    \label{fig:advice_toxic}
    \Description{Stacked bar chart of expert rankings of toxic content advice, advice shown in decreasing order of priority. 83\% of experts rated ``Mute people who post abusive messages'' as high priority, and 17\% rated as medium priority. ``Use platform-provided tools to automatically filter or moderate abusive messages'' was ranked second highest overall, followed by ``Block people who post abusive messages'' and ``Be selective about which online communities you participate in.'' Only 42\% of experts rated ``Be selective about when and to whom you reveal marginalized aspects of your identity'' as high, with 29\% ranking as medium and 20\% as don't recommend. ``Leave a platform entirely'' was the lowest overall, with 42\% ranking low and 25\% not recommending it at all.}
\end{figure*}

Experts ranked each of the 45 pieces of advice we collected  as ``high,'' ``medium,'' or ``low'' priority, or advice they ``don't recommend.''
In reasoning aloud, experts weighed factors such as efficacy, ease of implementation (and the existence of appropriate tooling), and whether advice curtailed a user's participation online. 
In this section, we review advice for staying safer from each threat, ordered by the average ranking of each threat from the prior section. We highlight only the advice that experts ranked highly, or where experts felt challenges persist or alternative solutions are needed. The complete set of advice is shown in Figures~\ref{fig:advice_toxic}--\ref{fig:advice_overloading}.\footnote{A unified, ranked list of all advice is included in the supplementary material.}

\paragraph{Preventing toxic content: Agreement about muting and blocking, but challenges around curtailing personal expression} 
To combat toxic content, experts favored platform-assisted moderation, with 83\% highly prioritizing \advicemute and 71\% \adviceblock (Figure~\ref{fig:advice_toxic}). 
Experts prioritized muting over blocking because blocking is more visible to attackers, who might escalate attacks when they find out they have been blocked.
Additionally, blocking impedes potential targets from monitoring their attackers:

\pquote{P4}{[Targets] don't want to read misogynist or racist comments, but they need to know that certain conversations exist, or whether they face threats. So they want to mute.}

{\noindent}Muting allows a target to quietly filter offensive users they encounter online (\eg community members), whereas ``blocking sends a signal you no longer want to interact'' (P24).
As such, experts noted that being aware of and being quick to use these features could curb future harm, in addition to their conventional use when there is an active attacker.

When asked if any advice to help prevent toxic content was missing, 13 experts said that reporting hate and harassment should be included,\footnote{During our advice gathering, we came across reporting, but at the time, we regarded it as not being proactive and thus out of scope for this study. However, we include it here because so many experts mentioned the importance of being aware of this feature. Additionally, reporting, like blocking and muting, are features general internet users should be aware of in advance, so they are prepared if or when attacks occur.} grouping it with blocking or muting as a standard best practice. 
Experts recommended reporting to the platform as well as to civil society organizations that can organize multiple reports, noting that reporting was a primary mechanism for platforms to find new issues and make improvements.
At the same time, experts lamented that ``reporting doesn't have an immediate impact'' (P16) and could be detrimental emotionally if the platform ultimately determined the reported attack did not cross a policy line:

\pquote{P24}{It's more harmful for the person [who submitted the report] to get a message that this wasn't even [determined to be] harmful.} 

While experts broadly agreed on the high prioritization of advice for mitigating toxic content, advice that required a user to limit their participation online was far more contentious, even when it was considered to be effective at preventing an attack.
Of experts, 63\% highly prioritized \adviceselectivecommunity and just 42\% \adviceselectiveaspect, while 29\% of experts did not recommend the latter at all. Among experts who rated either highly, a common refrain was being aware of unsafe communities and what you share as part of dealing with the realities of hate and harassment today:

\pquote{P15}{As a user, you should be able to decide... where you feel comfortable the most. If you don't feel comfortable on say, [platform], because a) you're not sharing that much and b) you're getting a lot of information pollution, or you don't find it useful at all, it makes sense to be selective.} 

\pquote{P20}{Heartbreaking. The whole idea of not being able to bring your whole self to an experience... Sadly I would always give that advice for today. I hope it's not advice I need to give in the future.}

{\noindent}Experts who were opposed expressed concerns that such advice required more nuance than was possible for a general guide. Others felt such recommendations gave up the ability to participate freely: 

\pquote{P10}{I understand the practical reasons behind it, but philosophically it's not right to expect people to do that... I've been doing stuff with [platform type], and there's this general philosophy we're trying to disrupt: `If you don't like it you can go somewhere else.' I don't like that sensibility being recommended from the top down.}

The most contentious advice for combating toxic content was \adviceleave. Only 13\% of experts ranked it highly, while 67\% put it as low priority or not recommended. Experts in support highlighted it could be appropriate as a last resort:

\pquote{P13}{It's always a tradeoff between having fun and not receiving too much harm... It's not the first thing you should do to deal with harm, you should try other things first. But if the harm is too pervasive and this is the only way to prevent it, they should.}

{\noindent}However, most experts opposed this advice due to losing voices of people targeted by hate and harassment, or the quality of life for following it:

\pquote{P10}{Just imagining the life of a perfectly secure user is really depressing. Is that really a life at all?} 

{\noindent}Experts recommended an alternative: taking a break or turning off notifications in order to disconnect. Broadly, advice for combating toxic content was more sparse compared to other threats we discuss. However, it was also one of the few threats with protections built-in to most platforms today.

\begin{figure*}[t]
\centering
\includegraphics[width=\textwidth]{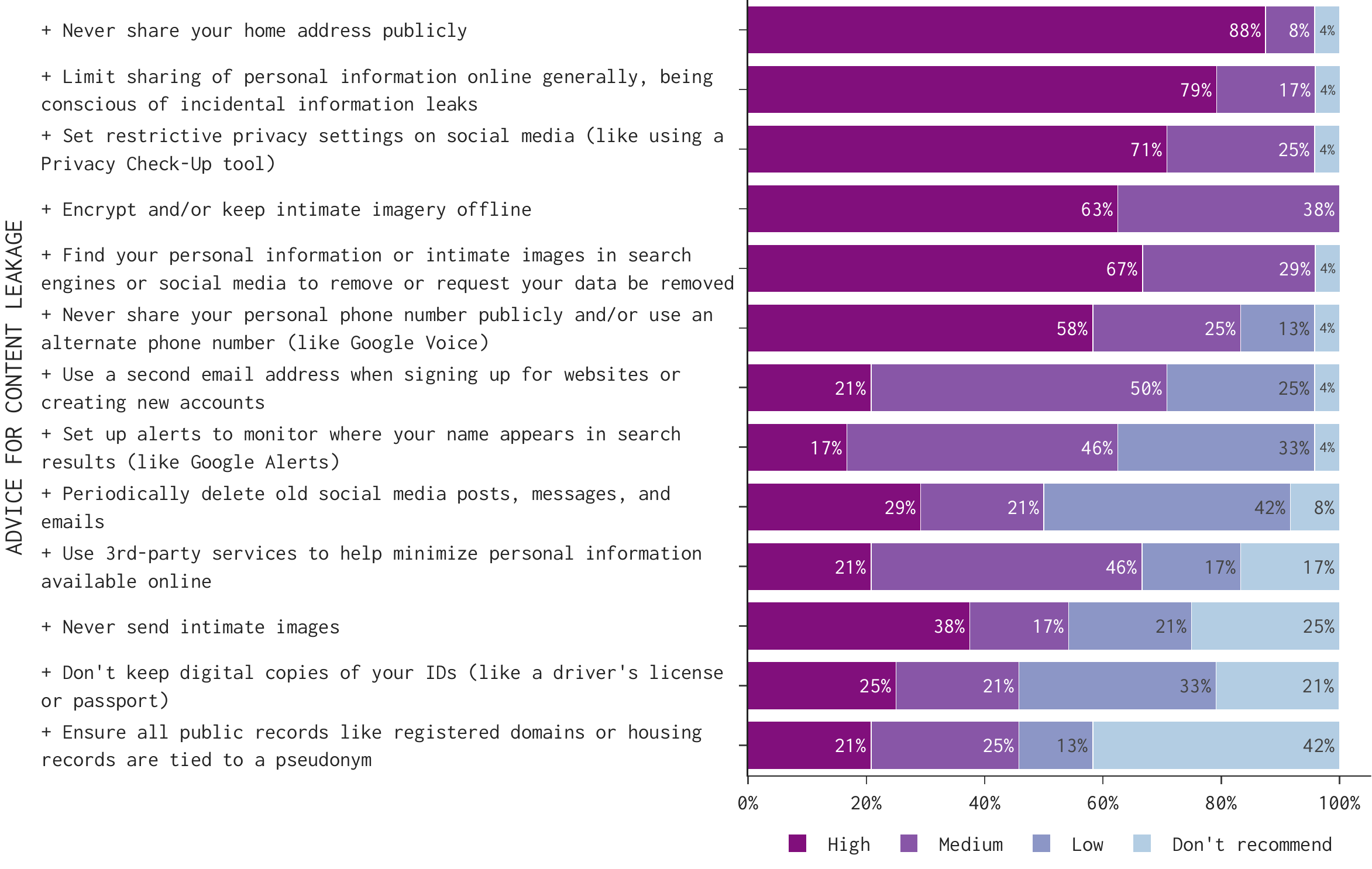}
\caption{Ranking of advice users could employ to help prevent content leakage. Experts prioritized advice involving data minimization involving one's address, phone numbers, and  personal information.}
\label{fig:advice_content_leakage}
\Description{Stacked bar chart of expert rankings of content leakage advice, advice shown in decreasing order of priority. Six pieces of advice were rated high priority by at least 58\% of experts: ``Never share your home address publicly,'' ``Limit sharing of personal information online generally, being  conscious of incidental information leaks,'' ``Set restrictive privacy settings on social media (like using a Privacy Check-Up tool),'' ``Encrypt and/or keep intimate imagery offline,'' ``Find your personal information or intimate images in search    engines or social media to remove or request your data be removed,'' ``Find your personal information or intimate images in search    engines or social media to remove or request your data be removed.'' The remaining pieces of advice were not ranked highly by more than 38\% of experts, tending to have a mix of of all rankings: ``Use a second email address when signing up for websites or creating new accounts,'' ``Set up alerts to monitor where your name appears in search results (like Google Alerts),'' ``Periodically delete old social media posts, messages, and emails,'' ``Use 3rd-party services to help minimize personal information available online,'' ``Never send intimate images,'' ``Don't keep digital copies of your IDs (like a driver's license or passport),'' ``Ensure all public records like registered domains or housing records are tied to a pseudonym.'' The last piece of advice was not recommended by 42\% of experts.}
\end{figure*}

\paragraph{Preventing content leakage: Agreement about the need to restrict information that's publicly available, but challenges with the ease of implementation and curtailing personal expression} 
To combat content leakage, experts recommended that individuals focus on restricting what information they share (Figure~\ref{fig:advice_content_leakage}).
88\% of experts highly prioritized \adviceshareaddress and 79\% highly prioritized \advicelimit, reasoning that ``the more information that's out there, the more potential for leakage'' (P11).
For other highly recommended advice, such as \advicesetprivacy, experts believed user awareness to be low: P3 commented that ``most people don't know they can change their settings.''

Though restricting information sharing was perceived as effective, experts discussed challenges with a cluster of advice that would be effortful to implement. 
For example, 58\% of experts highly prioritized not sharing personal phone numbers, but P6 noted that people might do so accidentally---``maybe you didn't intend to share it publicly but it's attached to a review or something.''
Similarly, only 25\% of experts reported that not keeping digital copies of IDs was a high priority, because digital copies of IDs are becoming very common and sometimes obligatory (\eg vaccination records to help manage the COVID-19 pandemic).
Other pieces of advice that experts thought could be helpful but would require excessive effort for a general internet user included using a second email address for accounts, using third party services to remove information online (\eg DeleteMe), or ensuring that public records like domain name registration or housing records are tied to a pseudonym.

Experts were very divided whether \adviceneversend should be recommended to prevent content leakage: 38\% prioritized it highly, 38\% prioritized it as medium or low, and 25\% would not recommend it.
Some experts noted that never sharing would be highly effective---``that's one of the easy ones'' (P12)---while other experts considered the advice to be victim blaming:
\pquote{P8}{If people want to share intimate images, technology should support their ability to do so.}
To sidestep issues of personal digital expression, experts were in greater agreement that people should \adviceencryptintimate, as 63\% highly prioritized doing so. Experts emphasized the offline part most---``don't use cloud storage'' (P7), ``prefer offline to encrypted'' (P3)---but mentioned ``there are a lot of tools now to keep these under lock and key'' (P24).
Experts also recommended other tips for sending intimate images more safely, such as only sending them to highly trusted people, or ensuring the images do not include identifying details such as one's face or tattoos.

Another challenge that experts noted for preventing content leakage was that certain pieces of advice would be relevant only for a subset of users.
Only 17\% of experts advised general users to \advicealerts:
\pquote{P20}{Only if you have some higher risk factor. Are you a streamer, or do you work in an industry where you deal with the public in a way that you are more likely to encounter harassment? Working at [a high profile company], this was a huge concern of mine.}
Other experts added that alerts were also only useful for people with unique names, and cautioned that alerts would lead to frequent false alarms for people with common names.

Similarly, experts judged that reviewing old content was only worth the effort for certain groups:
\pquote{P4}{People will go after you if you are a journalist and write about sensitive topics like politics or extremism. So they will search for what you wrote as a student from 10 years ago, which you may have forgotten about.}
67\% of experts considered \advicerequestremoval high priority to do once in a while, though P6 cautioned that overemphasizing this advice ``can make people really paranoid'' and ``only gives this advice if there is a reason, like someone saw a picture of you online or you have an abusive ex.'' 

\begin{figure*}[t]
\centering
\includegraphics[width=\textwidth]{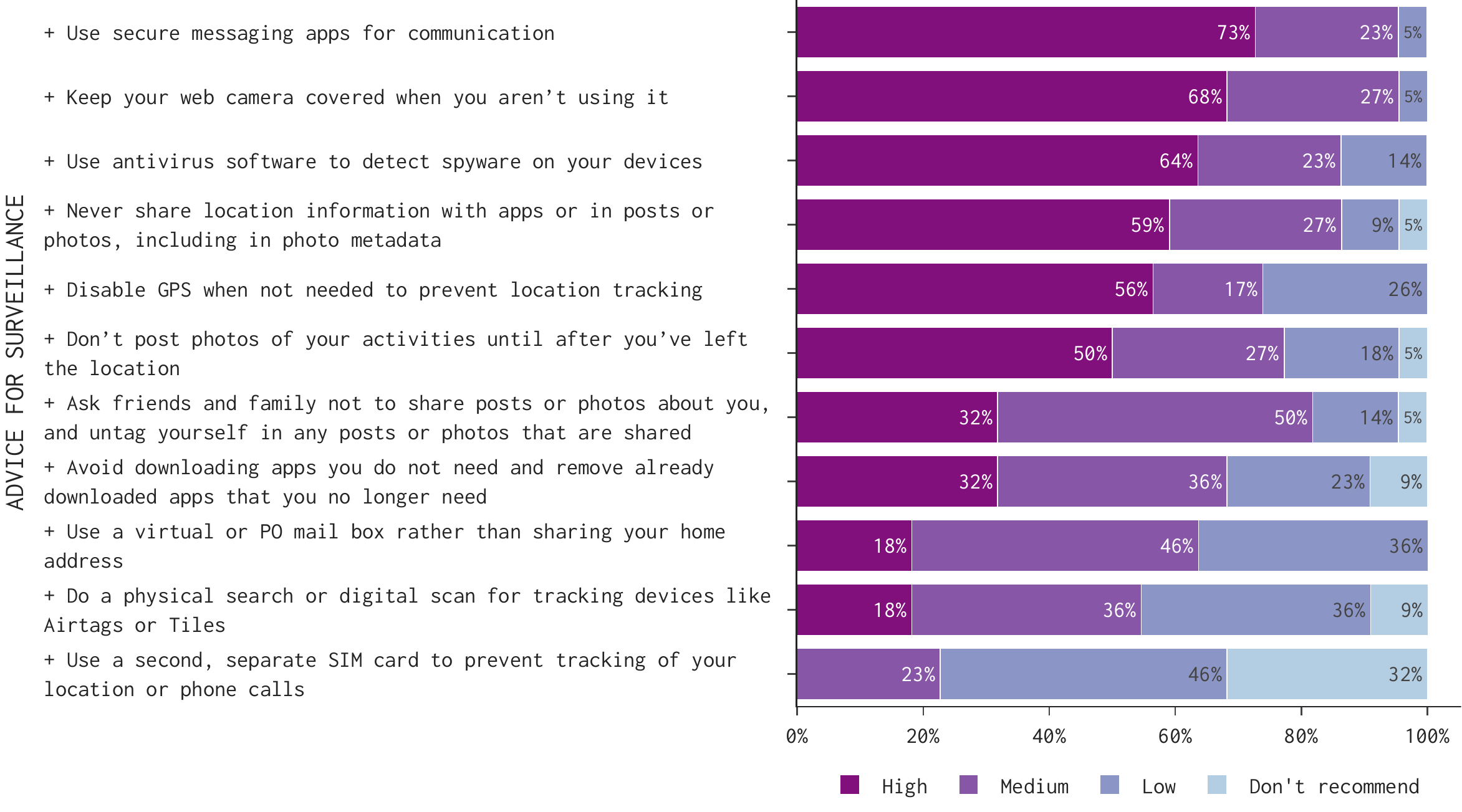}
    \caption{Ranking of advice users can employ to help protect themselves from surveillance. Experts prioritized making use of privacy tooling and limiting usage of certain application features.}
    \label{fig:advice_surveillance}
    \Description{Stacked bar chart of expert rankings of surveillance advice, advice shown in decreasing order of priority. Six pieces of advice were rated high priority by at least 50\% of experts: ``Use secure messaging apps for communication,'' ``Keep your web camera covered when you aren't using it ,'' ``Use antivirus software to detect spyware on your devices,'' ``Never share location information with apps or in posts or photos, including in photo metadata,'' ``Disable GPS when not needed to prevent location tracking,'' ``Don’t post photos of your activities until after you've left the location.'' Five were ranked high or medium by at least 50\% of experts: ``Ask friends and family not to share posts or photos about you, and untag yourself in any posts or photos that are shared,'' ``Avoid downloading apps you do not need and remove already downloaded apps that you no longer need,'' ``Use a virtual or PO mail box rather than sharing your home address,'' ``Do a physical search or digital scan for tracking devices like Airtags or Tiles.'' Finally, ``Use a second, separate SIM card to prevent tracking of your location or phone calls'' was ranked low or not recommended by 78\% of experts.}
\end{figure*}

\paragraph{Preventing surveillance: Agreement about the usage of privacy tools, but challenges around effectiveness and ease of implementation}
High priority advice for surveillance focused primarily on using strong privacy tools, or limiting certain application features that might leak one's location or identity (Figure~\ref{fig:advice_surveillance}).
However, experts' evaluation of advice surfaced challenges about whether advice would be effective in mitigating a surveillance threat such as stalking.

73\% of experts highly prioritized \advicesecuremsg, but multiple experts viewed secure messaging more through a lens of general security threats, rather than hate and harassment. For example, P16, who ranked the advice as high priority, explained: ``I do recommend [secure messaging] to people, maybe not in this [hate and harassment] context, but I generally do.'' 
Other highly ranked advice for mitigating surveillance via compromised devices was also more protective against general threats, and less aligned to surveillance for hate and harassment. Advice such as \advicecamera and \adviceav were highly prioritized by 68\% and 64\% of experts respectively, as they were seen as supporting user agency---they are simple steps that could provide some protection: ``no harm in doing it, but I wouldn't say you need to go home tonight and cover every web camera''  (P14).
Yet, P8 clarified that cameras were only a superficial concern for surveillance and ranked this as low priority:

\pquote{P8}{[You're] not dealing with the root cause. If you're worried about your web camera, [you] should be worried about bad software in general on your device.}

{\noindent}Thus, despite experts finding some advice in this section high priority, there remains room for new advice and protections that would more effectively protective against surveillance.

Experts were generally not in favor of other more strict physical access measures such as \advicepobox, \adviceairtags, or \advicesecondsim due to the substantial effort of implementing the advice. Experts felt this advice ``really depends on your threat model'' (P9) and expressed that they were ``not sure creating an atmosphere of anxiety is needed'' (P20) for general internet users. However, experts noted that in some contexts, these practices became critical:

\pquote{P11}{If you are running from an abusive spouse, then absolutely... But I wouldn't recommend everyone in the world do this. }

{\noindent}Experts also warned of the challenges of enacting this advice successfully. Searching for physical tracking devices is ``really difficult to do... people don't know how to do a digital scan" (P12) and ``may not be possible for people who aren't well versed'' (P15), echoing Gallardo \etal's findings that detecting surveillance issues is difficult~\cite{gallardo2022iphone}. Likewise, ``it's a lot of work to get a P.O. box for all deliveries. It's inconvenient for real life'' (P12). As a whole, experts felt this advice was best suited to people who knew they were in a surveillance situation, but not something that general internet users needed to be concerned about.

\begin{figure*}[t]
    \includegraphics[width=\textwidth]{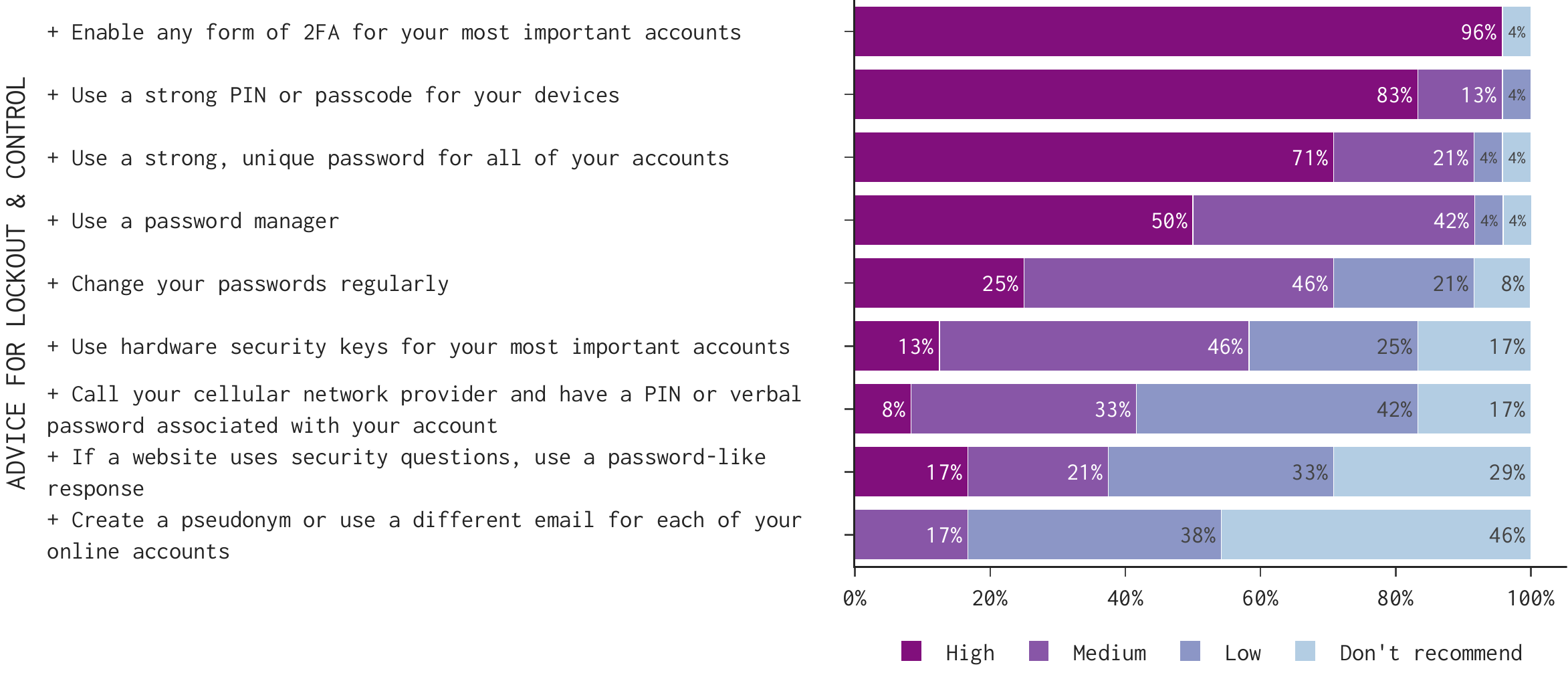}
    \vspace{-20pt}
    \caption{Ranking of advice users could employ to help prevent lockout and control. Experts limited their advice to proven account security best practices.}
    \label{fig:advice_lockout}
    \Description{Stacked bar chart of expert rankings of lockout and control advice, advice shown in decreasing order of priority. Four pieces of advice were rated high priority by at least 50\% of experts: ``Enable any form of 2FA for your most important accounts,'' ``Use a strong PIN or passcode for your devices,'' ``Use a strong, unique password for all of your accounts,'' ``Use a password manager. Two were ranked high or medium by at least 50\% of experts: ``Change your passwords regularly,'' ``Use hardware security keys for your most important accounts.'' Finally, three were ranked low or not recommended by at least 59\% of experts.``Call your cellular network provider and have a PIN or verbal   password associated with your account,'' ``If a website uses security questions, use a password-like response,'' ``Create a pseudonym or use a different email for each of your   online accounts.''}
\end{figure*}

\paragraph{Preventing lockout and control: Agreement about establishing account hygiene, but challenges with the ease of implementation}
To protect against account-based threats, experts overwhelmingly favored protections they considered to be basic account hygiene (Figure~\ref{fig:advice_lockout}). 96\% of experts highly prioritized \advicetwofa, as did 83\% \advicepin, and 74\% \advicestrongpwd. 
As P16 explained regarding 2FA:

\pquote{P16}{If you are actually worried about people hacking [your account], a password isn't enough.}

{\noindent}Experts also discussed how 2FA alleviates the need for users to change passwords regularly, noting the reality that many users do not use strong \textit{or} unique passwords. Experts also noted that users are becoming more familiar with it and finding it ``less horrible''~\cite{colnago20182fa} than they expected.
Only one expert did not recommend 2FA because ``people get locked out of basic services often'' (P12).

Favorability of 2FA stopped short of hardware keys (as opposed to SMS or on-device prompts), with just 13\% of experts stating hardware keys were a high priority, mainly because it was unnecessarily burdensome for general users.
P16 felt this level of security was only needed ``if you have the nuclear codes'' while others stated this was more important if you had business secrets or professional accounts that might be targeted.

The effort necessary to protect against attackers exploiting weak security questions or having multiple accounts to avoid a single source of failure was also viewed as too onerous. Of experts, 62\% rated \advicesecquestion and 84\% rated \adviceunlinkableaccts as low priority or not recommended. For hardening security responses, experts were concerned primarily with users forgetting responses.
For managing multiple accounts, experts felt the credentials would be too much to remember:

\pquote{P11}{How are you going to keep track? ...we've all got at least 10 or 20 different accounts.}

When asked about any missing advice, experts added four pieces for helping prevent lockout and control: keeping account recovery vectors up-to-date (mentioned by 2 experts), checking whether passwords have been exposed by a breach (2), never sharing passwords (1), and keeping an eye out for notifications of suspicious account logins (1).

\begin{figure*}[t]
\centering
\includegraphics[width=\textwidth]{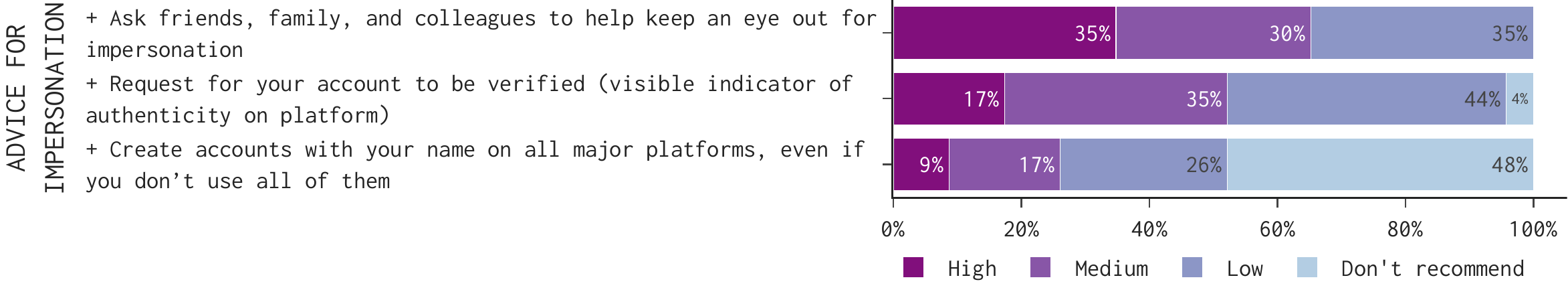}
    \caption{Ranking of advice users could employ to help prevent impersonation, none of which experts felt was effective for general internet users.}
    \label{fig:advice_impersonation}
    \Description{Stacked bar chart of expert rankings of impersonation advice, advice shown in decreasing order of priority. Two pieces of advice was rated high priority by at least 50\% of experts: ``Ask friends, family, and colleagues to help keep an eye out for impersonation,'' ``Request for your account to be verified (visible indicator of  authenticity on platform).'' Finally, one was ranked low or not recommended by 74\% of experts.``Create accounts with your name on all major platforms, even if you don't use all of them.''}
\end{figure*}

\paragraph{Preventing impersonation: Lack of effective advice} 
Across experts, there was no existing advice---nor any advice they could provide---that a consensus felt was high priority to help prevent impersonation (Figure~\ref{fig:advice_impersonation}). Advice such as \advicefamilyimpersonation were ranked as both high and low priority by 35\% of experts. As a proactive practice, most experts viewed this as too ``paranoid,'' particularly in light of the low prevalence of impersonation in their experiences. Similarly, experts raised concerns about feasibility. As P4 put it:

\pquote{P4}{Do you really think your friends and family and colleagues will spend the time to look out for impersonation for you? They don't care. They have so many things to do.}

{\noindent}Experts felt this advice was more pertinent when responding to an active or previous attack (\ie if someone has been or is being impersonated):

\pquote{P14}{If you were being targeted, you should do this. But not if you didn't have reason to believe you were being targeted.}

Experts also deemed other forms of bolstering one's digital identity as infeasible or ineffective: 48\% ranked \adviceverifyaccount as low priority or not recommended, while the same was true for 74\% of experts when ranking \adviceaccountsquatting. Verification (\eg a visual indicator of trust available on many social media platforms) was perceived as restricted by platforms to celebrity-like individuals who had a sufficiently large audience, and thus beyond the capabilities of most internet users.\footnote{Our interviews were conducted several months before the December 2022 roll out of \textit{Twitter Blue}.}
Likewise, managing multiple accounts that a user wasn't planning to actively use was viewed as burdensome and potentially even harmful due to compounding account security risks (\eg the reality that many users would likely use weak passwords). 

\pquote{P9}{I don't recommend that at all. That's basically saying you need to sign up for everything... If you don't have good password hygiene and use the same password on all of them, you can be compromised faster.}

The lack of advice for impersonation stems, in part, from the challenge that attacks frequently occur without a target's knowledge, and often on platforms where the target is not a participant (\eg fake dating profiles, fake social media accounts).

\begin{figure*}[t]
\centering
\includegraphics[width=\textwidth]{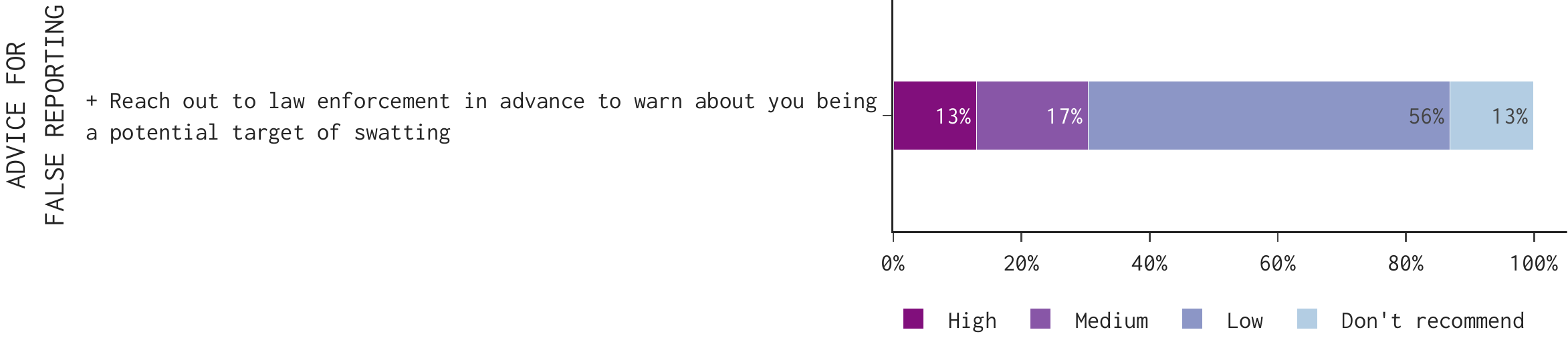}
    \caption{Ranking of advice users could employ to help prevent false reporting. Experts viewed swatting as outside the scope of general internet user threat models. Likewise, law enforcement might not be equipped to handle warnings.}
    \label{fig:advice_false_reporting}
    \Description{Stacked bar chart of expert rankings of false reporting advice, advice shown in decreasing order of priority. Only one piece of advice, ``Reach out to law enforcement in advance to warn about you being a potential target of swatting,'' which was ranked high by only 13\% of experts and low by 56\% of experts, with the rest split between medium or not recommended.}
\end{figure*}

\paragraph{Preventing false reporting: Lack of effective advice} 
When gathering existing advice, the only advice we found to combat false reporting was to \adviceswat (Figure~\ref{fig:advice_false_reporting}). A majority of experts---69\%---ranked this as either low priority or not recommended, most commonly because of the low prevalence of swatting on general internet users:

\pquote{P1}{If you’re likely to get swatted, then it’s a high priority. If you’re just a regular person and you did this, the police would think you're crazy ... In the general case, you shouldn't even think about [being swatted].}

{\noindent}Other concerns focused on the perceived indifference of law enforcement, a lack of law enforcement training on how to handle such warnings, or a general distrust of law enforcement (particularly in authoritarian regions):

\pquote{P20}{This one is complicated. A lot of times law enforcement isn't well set up to do anything with this information. Maybe a good idea, but it's contingent on where you are in the world.}

While swatting is the most severe form of false reporting in terms of physical harm, there remains a lack of helpful advice for attacks that attempt to silence a target by having their account terminated. Such attacks depend entirely on the procedures and practices of third-party platforms, which targets can only partially navigate by choosing where they participate.

\begin{figure*}[t]
\centering
\includegraphics[width=\textwidth]{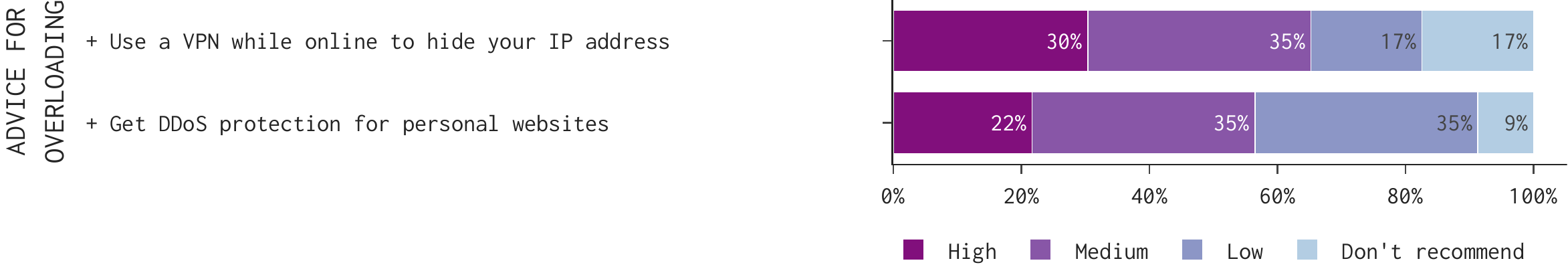}
    \caption{Ranking of advice users could employ to prevent  overloading. While VPN and DDoS protection services exist, experts felt they were too cumbersome or out-of-scope for most hate and harassment that general internet users would experience.}
    \label{fig:advice_overloading}
    \Description{Stacked bar chart of expert rankings of overloading advice, advice shown in decreasing order of priority. Only two pieces of advice, ``Use a VPN while online to hide your IP address'' and ``Get DDoS protection for personal websites'' which were both ranked by about 60\% of experts as high or medium priority.}
\end{figure*}

\paragraph{Preventing overloading: Lack of effective advice} 
While overloading encompasses multiple threats---such as notification bombing, brigading, or dogpiling---existing online advice we found was limited solely to network security (Figure~\ref{fig:advice_overloading}). For \advicevpn, there was a large spread of prioritization among experts. For P15, this was a ``general thing that everyone should be doing,'' whereas for P8, this advice was ``pretty in the weeds and not relevant to most, but if you're targeted, could be reasonable.'' Other concerns included barriers to access, usability concerns around proper configuration, and misconceptions about what protections VPNs provide (as recent work has also explored~\cite{binkhorst2022vpn, ramesh2023VPN, akgul2022VPN}). 

Similarly, \adviceddos was prioritized as either medium or low by 70\% of experts. P22 felt it was a ``no brainer, but not easy,'' whereas most experts felt this advice should be restricted to people who had personal websites with a higher likelihood of being targeted. 

The lack of guidance for brigaiding or dogpiling---such as when a person goes viral outside their intended audience---exposes a critical gap in advice today for general internet users. This is particularly problematic as these attacks occur spontaneously, limiting the window for a target to react, or to control the spread of their content once its shared beyond spheres where they have platform-provided privacy controls.

%% file: 06_results_strategies.tex
\subsection{Overall Safety Strategies}\label{ss:advice-top}
When we asked experts to describe their personal top three recommendations for general internet users with respect to online hate and harassment, we received responses that varied greatly in specificity.
Some experts named discrete actions, such as pieces of advice from Section~\ref{ss:advice-ranking}, while others spoke broadly about things users should keep in mind.
We synthesize the 65 top recommendations of the experts we interviewed below.\footnote{Most, but not all, experts gave top recommendations. One expert passed on giving any top recommendations, explaining that one-size-fits-all advice did not exist. Some experts combined multiple recommendations, so counts do not sum to 65.}

\paragraph{Data Minimization (recommended 24 times)}
Across all experts, the most common top recommendation was to minimize sharing personal information. Experts spoke about the importance of reducing the amount of personal information that is available online, both by being mindful of what a user shares, as well as deleting existing data that is already online.
However, experts were also cautious about recommending that people limit what they share online noting that it ``may not eliminate the potential for things to happen'' (P23).
Going further, P23 explained that data minimization is not a sustainable solution:
\pquote{P23}{Putting limits on self-expression may keep you safe in the short term but it's not good for the health of online spaces overall.} 
Echoing this concern, P8 reasoned that the framing of the advice would be crucial:
\pquote{P8}{Being careful about what you put online is always a reasonable thing to suggest to people. It is a little victim-blaming at the end of the day, right? So it has to be worded appropriately, but certainly good advice.}

In addition to limiting sharing, experts favored auditing security and privacy settings, especially for social media accounts or location tracking.
P24 noted that it was important to consider how information is presented online, and making sure that users know who content is visible to.
Privacy and security settings, similar to limiting information available, were seen by experts as actions where users had agency, which may be why they were the most common pieces of top advice.
Further, these recommendations align with our finding that content leakage was, on average, the second most important hate and harassment threat that experts thought general users should be concerned with (see Section~\ref{ss:threat-ranking}).

\paragraph{Account Security (recommended 18 times)}
Experts frequently recommended general account security practices, including using 2FA, creating strong and/or unique passwords, and using a password manager.
P3 described these tips as putting yourself on the path of least resistance: 
\pquote{P3}{You don't have to set up the most complicated security system you can think of. Do things that will slightly deter you from having a bad experience online compared to the general public.}

\paragraph{Self-Determination and Awareness (recommended 17 times)}
Experts believed that users should determine for themselves \textit{where} they choose to engage online:
\pquote{P2}{Consider the community you're engaging in and its culture... if you're going to be on 4chan, you're going to get hateful content... so it's better to start off in more protected, smaller, or closed communities with better norms.}
By being more aware of the community norms, as well as the potential protections afforded by certain platforms, experts reasoned that users could better avoid harm.
Experts also recommended that users pay attention to \textit{how long} to engage online, or in P2's words, ``decide for yourself how much bullying or harassment you're willing to endure.'' 
By determining how much abuse an individual is willing to tolerate, experts reasoned that users could decide when to ``leave the platform, especially if it's continuous and targeted -- the platform isn't for you'' (P11) or at least temporarily ``remove yourself from any situation from which you feel unsafe'' (P20).

In a similar vein, experts recommended that users stay aware of how they might be threatened, and what existing tools could help.
Searching for yourself online was seen as a good way to ``be aware in general of your digital footprint or online presence'' (P15).
Given that threat modeling is a standard practice in security for enumerating threats, two experts explicitly recommended it, and one expert implicitly: ``Think deeply about who has access to your devices and how you keep those secure'' (P24).

\paragraph{Safer Through Community (recommended 9 times)}
The final strategies recommended by experts were communally-focused.
Experts recommended reporting hateful or harassing content---``my favorite is still: block aggressively'' (P7)---not only for immediate individual relief, but also because doing so would ultimately help foster safer online communities.
\pquote{P18}{Don't be a silent bystander... we're not going to create a better world by being silent about it. Use the tools you've got. If you can report, report. If you can stand up for folks, stand up for folks... So it's not just about protecting yourself, it's about being a good digital citizen. It's important because if you're waiting for others to change, there won't be change.}
Other experts further supported the need for pro-social behaviors that would improve broader online communities by proactively looking out for others, as well as sharing the responsibility for creating healthier online environments. If users do experience harm, one expert recommended reaching out for help from trusted parties.
P13 hoped people who have been targeted would understand that: 
\pquote{P13}{It's not your fault. As long as we expose ourselves online, there are dangers that we face. Many times, survivors blame themselves for it. They aren't sure whether it's harm or if they're overreacting. Or they think that they did something wrong so they should be blamed for receiving harassment. The internet environment can be toxic sometimes, and platforms may have given you limited tools to address the harassment, so you feel like you have less agency, but it's not your fault. We should acknowledge that others have responsibility to protect them.}

%% file: 07_discussion.tex
\section{Discussion}\label{sec:discussion}

In this work, we sought to find generally applicable advice that would contribute to individual safety from online hate and harassment without additional context about the user.
From an interview study with subject matter experts, we outlined a cluster of top threats they believe users should prioritize and advice users can employ to help prevent those threats, as well as overall safety strategies.
We now step back to discuss tensions our work surfaces for efforts to help people stay safer from online hate and harassment.

\paragraph{Our work illustrates the complementary roles of general and tailored advice}
Though our aim was to explore general advice, the current landscape of online hate and harassment makes both general and tailored advice valuable, given the unique benefits and limitations of each.

Most prior hate and harassment safety advice---including the advice we collected for our work---takes a tailored approach.
Tailored advice centers marginalized populations that are at disproportionate risk for online hate and harassment, providing  invaluable support to those who may need it the most.
Yet, tailored advice is extremely challenging to create and maintain.
Experts in our study who served as advocates for specific populations expressed that existing resources were insufficient, despite not even serving all groups that need support. 
Further, groups needing tailored support may not know such resources exist or how to find them.
Therefore, tailored advice is best for users who understand that they are at a disproportionate risk and helps them focus their effort where it will be most effective.

Contrasting tailored approaches, prior work on traditional security and privacy advice has called to ``identify the smallest and most easily actionable set of behaviors to provide the maximum user protection''~\cite{redmiles2020comprehensive}.
In some contexts, such as when advice-givers do not have more detailed information about users' situations or when users do not wish to reveal sensitive information about their situation, general advice is the only viable option.
General advice empowers individuals to adopt effective safety practices with lasting consequences even before they are at risk
or become aware of tailored advice for their situation. 
Users are also more likely to follow general advice that multiple sources consistently repeat, though such advice approximates an average threat level and can under- or over-prepare potential targets.
Therefore, general advice is best viewed as a baseline of protection for a wide range of users, and as a stepping stone towards tailored advice.

Throughout this work, we grappled with the need for advice that would be relevant for an ever-increasing proportion of internet users who will face online hate and harassment and the heterogeneous experiences that each user will have. 
Both general and tailored advice can have a valuable role in supporting potential targets.
Our study further shows that general advice rarely contradicts tailored advice; instead, general advice is best for when less information about users is available, and tailored for when more information is.

\paragraph{The lack of consensus on top threats poses a challenge for which education and safety tools advocates should focus limited resources towards developing}
In the absence of contextual information about a person's unique needs, experts only loosely agreed on which threats general internet users should prioritize preventing or mitigating. Part of this complexity stemmed from the three competing dimensions that experts used to rank threats: severity of harm from the threat, prevalence of the threat, and agency that users have to combat the threat. 
For example, some experts who had experience supporting targets of intimate partner abuse were especially attuned to the \textit{severity} of threats posed by targets' intimate partners, ranking lockout and control as well as impersonation threats higher than other experts.
But some experts who supported journalists or content creators whose jobs necessitated they have a prominent online presence were particularly attuned to \textit{prevalent} forms of online hate and harassment, tending to rank toxic content higher.
Our interviews did not indicate a clear path for resolving these tensions, or if such a path even exists. 
Experts also leaned on their considerable, deep experience for the particular populations they served, which do not represent all people who experience hate and harassment. A remaining question for future work is: how might research and practice deliver relevant advice to people's unique risk profiles \textit{at scale}, especially if particular at-risk groups are not yet understood?

While better empirical measurement may assist arriving at a consensus and thus how to best allocate resources, some applications might necessitate prioritizing one dimension over others. Companies with broad user bases might focus on prevalence, acknowledging that severity and agency fall to other actors. Specialized support providers, such as for survivors of intimate partner abuse, might center their efforts on high-severity threats. 
Taken together, these efforts would aim to communally balance the needs of specific groups that are at heightened risk for specific types of hate and harassment, while also considering some other users may never face such risks.
The multiplicitous approach also addresses a caution from prior work ``against using worst-case scenarios when average-case is what users care about'' ~\cite{herley2014more}. The average case of hate and harassment is not yet known and could very well change over time. Further, the nature of hate and harassment incidents does not allow for clean distinctions between ``average'' and ``worst.''

\paragraph{Effective advice requires letting a user make their own decision, at the right moment}
Many experts emphasized that how and when advice is offered is challenging, if not more so, than developing the advice itself. Our evaluation of advice centered which practices would be most helpful, and was less concerned with the particular phrasing, given different platform features (\eg restricting vs. muting accounts).
Further, many experts criticized the wording of advice that was prescriptive, explaining that starting advice with ``never'' (\eg never share intimate images) could be a non-starter. Instead, P8 described that allowing users to decide for themselves whether to adopt such advice would improve adoption, by ensuring they fully understood the protections and trade-offs of a given piece of advice. This sentiment echoes prior work on security behaviors broadly: ``that the benefit [of following security advice] is greater than the cost must be shown, not assumed or asserted''~\cite{herley2014more}.
This further embodies the principle of  \textit{enablement} from trauma-informed computing (which builds on the premise that accounting for trauma's effects is widely beneficial for all users, traumatized or not): computing should enable users to make informed decisions for themselves~\cite{chen2022trauma}. 

As with other security advice, experts pointed to times when people might be more receptive to enacting advice, such as after personal experiences with hate and harassment, or after hearing about others' experiences. However, delaying the adoption of advice until after an attack occurs may expose the target to irreversible harms (\eg content leakage). Such complexities reiterate the need for proactive advice that is generally applicable in the absence of knowing which threat might occur, complementing crisis resources to provide redress after a harm has occurred. 

\paragraph{The (apparent) effectiveness of some advice is at tension with the tendency for such advice to further perpetuate and entrench marginalization}
Expert opinion was divided on advice seen as effective that also significantly curtailed personal expression (\eg \adviceneversend, \adviceselectivecommunity, \adviceselectiveaspect).
Some experts judged this advice to be up to personal decision, so users have the final say on what they are comfortable with.
However, other experts highlighted how certain advice might seem effective now, but also systematically problematic.
For example, never sending intimate images could make content leakage less likely, but it may be interpreted as implying that those who initially send intimate images are at fault and not the perpetrators who actually leak (\ie nonconsensually share) such content.
P8 commented that such advice was victim-blaming because technology should support users in how they choose to express themselves online.
Further, self-limiting advice entrenches the marginalization that certain populations already endure.
Experts described that some gamers who are women and/or Black avoid harassment by not joining voice channels with strangers, at the expense of their own enjoyment of the games. 

Experts discussed the ways that the burden of avoiding harassment online is inequitably distributed, with marginalized populations having already accepted limitations to self-expression in order to exist online.
Yet, when experts described how advice for at-risk populations---such as journalists, survivors of intimate partner abuse, or content creators---might differ from general internet users, there was a tendency to strictly recommend more advice, in addition to other high priority advice for all.
This poses an untenable burden for marginalized groups to enact tens of pieces of advice for each type of threat.
As prior work has stressed, ``spending more time on security is not an inherent good''~\cite{herley2014more}.

\paragraph{The status quo places greatest responsibility on individuals to keep themselves safe, necessitating new solutions}
Many experts remarked that a majority, if not all, of the burden for staying safer online currently fell to users, reiterating a prior observation that ``we [the HCI \& security communities] have used user effort as a first resort, not last''~\cite{herley2014more}.
In order to reduce the need for individual responsibility,
many experts commented on the larger need for building communities with norms against hate and harassment.
Additionally, experts pointed to the benefits of social support networks in coping (\eg identifying friends who can provide emotional support) if online hate and harassment occurs.
In one expert's estimation, reassurance was a large portion of support:
\pquote{P15}{More than anything, people need comforting, someone to tell them that they’re okay.}
Social support might be especially valuable for threats where individual agency is low, and thus advice is sparse. For example, there was more advice for content leakage where privacy controls were a central defense, versus overloading or false reporting where attacks depended heavily on attacker capabilities and third-party practices.

These directions work in tandem with producing general and tailored advice. 
Advice serves as a critical, interim protection during the process of systemic change. 
Through both individual and communal effort, we hope to create a safer internet for all.

%% file: 08_conclusion.tex
\section{Conclusion}\label{sec:conclusion}

In this work, we conducted interviews with 24 subject matter experts to understand which pieces of advice can broadly and immediately help most internet users stay safer from online hate and harassment.
We used a lens of security and privacy to tackle the broad online hazard of hate and harassment, decomposing it into a set of technology-mediated threats to develop pragmatic guidance for anyone who might be a potential target.
Experts weighed different criteria to determine which threats should be prioritized, \ie prevalence or (potential) severity of the threat, as well as individual agency. This resulted in an overall ranking of toxic content, content leakage, and surveillance as the top three hate and harassment threats most internet users should take action to prevent or mitigate.
Further, we note the factors experts used to evaluate existing pieces of advice---efficacy, ease of implementation, and effect on online participation---and find a select few pieces of advice experts agreed were broadly applicable, while many other threats lacked suitable advice for users to implement.
Overall, our work identifies technical and design directions to support users in staying safer from online hate and harassment, while surfacing tensions and challenges on the notion of individual responsibility to do so at all.

%% file: acks.tex
\begin{acks}
We are deeply grateful for and recognize the contributions of our expert participants: Eve Crevoshay, Molly Dragiewicz, Jennifer Golbeck, Arzu Geybulla, Weszt Hart, Laura Higgins, Caroline Humer, Rachel Kowert, Liz Lee, Kat Lo, Thomas Ristenpart, Linda Steiner, Gianluca Stringhini, Leonie Tanczer, Elodie Vialle, Viktorya Vilk, Jessica Vitak, Kimberly Voll, Daricia Wilkinson, Sijia Xiao, and our anonymous experts.
We thank our reviewers for their valuable suggestions in improving our paper, Anna Turner and Stephan Somogyi for piloting our study, and Tara Matthews for reviewing our methods.
This work was supported in part by the U.S. National Science Foundation under award CNS-2205171 and a gift from Google. 

\end{acks}